# THERMAL EQUILIBRIA OF ACCRETION DISKS

Marek A. Abramowicz[1,2], Xingming Chen[1], Shoji Kato[1,3] Jean-Pierre Lasota[4,1], and Oded Regev[2]

[1]Department of Astronomy and Astrophysics, Göteborg University
and Chalmers University of Technology, 412 96 Göteborg, Sweden

[2]Department of Physics, Technion, Haifa 32000, Israel

[3]Department of Astronomy, Kyoto University, Sakyo-ku, Kyoto 606-01, Japan

[4]UPR 176 du CNRS; DARC, Observatoire de Paris, Section de Meudon, 92195 Meudon Cedex, France

## ABSTRACT

We show that most of hot, optically thin accretion disk models which ignore advective cooling are not self-consistent. We have found new types of optically thin disk solutions where cooling is dominated by radial advection of heat. These new solutions are thermally and viscously stable.

*Subject headings:* accretion, accretion disks — instabilities

# 1. INTRODUCTION

In the standard approach to stationary accretion disk models (Shakura & Sunyaev 1973) the energy equation takes into account only local, usually radiative, cooling that balances the viscous heating. In most cases global heat transport is neglected. This approach is justified as long as the disk is geometrically thin because the ratio of the advection term representing the global entropy transport to the viscous heating term is of the second order in the relative disk thickness $H/R$, where $H$ and $R$ are the semi-thickness of the disk and the distance from the central object respectively. In some time-dependent calculations of thin models (e.g. Taam & Lin 1984; Lasota & Pelat 1991) the advective heat transport is taken into account but it represents only a small contribution to the total energy balance. In the slim disk approach (Abramowicz, Lasota & Xu 1986; Abramowicz et al 1988; Kato, Honma, & Matsumoto 1988; Chen & Taam 1993) terms up to the second order in $H/R$ are taken into account so that the advection is part of the scheme together with the $(H/R)^2$ – order terms in the radial Euler equation. In the high accretion rate regime, when $H/R$ is not negligibly small, the cooling of optically thick slim disks is always dominated by advection. In the radiation–pressure dominated slim disk models the advective cooling rate per unit surface $Q_{adv}$ satisfies the relation (Abramowicz et al 1986)

$$Q_{adv} \gtrsim \left(\frac{H}{R}\right)^2 Q_+, \qquad (1)$$

where $Q_+$ is the viscous heating rate per unit area.

Recently Narayan & Yi (1994) constructed self-similar, advection dominated accretion flows with $Q_{adv} = Q_+$ and with a sub-Keplerian angular momentum distribution.

In the last few years, several hot optically thin disk models have been constructed following theoretical developments in the description of the microphysical processes which may be relevant for the cooling (see e.g., Svensson 1982, 1984). Attention has been focused on the electron-positron pair production and annihilation (Kusunose & Takahara 1988, 1989, 1990; Tritz & Tsuruta 1989; White & Lightman 1989, 1990; Björnsson & Svensson 1991a,b, 1992; Kusunose & Mineshige 1992). On the other hand, the relation between optically thick and thin disks has been studied by Wandel & Liang (1991) and Luo & Liang (1994) with purely phenomenological "bridging formulas" for the case of effective optical depth around unity. The latter authors have concluded that optically thick and thin sequences of solutions are always connected in the $\dot M$ (mass accretion rate) versus $\Sigma$ (surface density) plane.

In all those models the hot inner regions of the accretion disk are not geometrically thin because of the relative inefficiency of the optically thin cooling which makes $H/R \approx C_s/v_K$ close to unity ( $C_s$ is the adiabatic speed of sound and $v_K$ the Keplerian speed). Advection is neglected. Furthermore, they are all thermally unstable. This is not just a technical difficulty that could be repaired without modifying the basic properties of the proposed models. In most cases, the a posteriori calculated advective heat transport is found to be the *dominant* cooling mechanism at high accretion rates. As an example, we show in Figure 1 the equilibrium curves on the $(\dot M, \Sigma)$–plane for various models from Kusunose & Mineshige (1992) and Luo & Liang (1994); together with the advection dominated lines calculated with equation (10) modified to meet their assumptions and units. In Figure 1a, advection is dominant above the heavy dotted line; the heavy dashed line corresponds to $H/R = 1$ (see eq. [7]). Note that equation (10) corresponds to an advection-dominated solution independent of the local cooling mechanism (see also Fig. 3). One can see that, most of the high temperature solutions are located in the regime where advective cooling is dominant. Since it was neglected, the presented models are *physically* inconsistent. Note also that the geometrically thin disk assumption is not satisfied either. This is less serious because solutions inconsistent in this sense often possess all qualitative properties of the real configurations. In Figure 1b, the advection dominant lines are represented by the solid, dotted, and dashed ones corresponding to $\alpha = 1$, 0.1, and 0.01 respectively. It is seen that for $\alpha \lesssim 0.1$, the models are not self-consistent.

## 2. ADVECTION DOMINATED OPTICALLY THIN SOLUTION

These difficulties have their source in neglecting the dominant effect of the advective cooling. Thermally stable optically thin accretion disk models with $H/R < 1$ can be constructed when this effect is included. One can get an insight into the properties of optically thin advection – dominated accretion disks by assuming that advection and other (local) mechanisms cool a *Keplerian disk*.

We thus assume the rotational angular velocity is $\Omega = \Omega_k = \sqrt{GM/R(R-R_G)^2}$ under the pseudo-Newtonian potential $\Phi = -GM/(R-R_G)$, where $M$ is the mass of the central object and $R_G$ is the Schwarzschild radius (Paczyński & Wiita 1980). Since the optical depth is equal to zero the disk is gas–pressure dominated, $P = P_g = \mathcal{R}\rho T/\mu$. Here $\rho$ and $T$ are the density and the temperature of the disk respectively, and $\mu$ is the mean molecular weight assumed to be 0.617. Since we are interested in hot disks, opacity will be given by electron scattering: $\kappa_{es} = 0.34$. We shall assume that radiative cooling is provided by optically thin thermal bremmstrahlung with emissivity per unit area,

$$Q_{brem} = 1.24 \times 10^{21} H\rho^2 T^{1/2} \text{erg s}^{-1} \text{ cm}^{-2}. \qquad (2)$$

The viscous heating rate per unit area is given by the standard formula

$$Q_+ = \frac{3G}{4\pi}\frac{M\dot{M}}{R^3}fg, \qquad (3)$$

where $\dot{M}$ is the mass accretion rate, $g = -(2/3)(d\ln\Omega/d\ln R)$, and the factor $f = 1 - \Omega(3R_G)/\Omega(R)(3R_G/R)^2$ contains the inner boundary conditions. The advection cooling rate is taken in a form (see e.g. Chen & Taam 1993):

$$Q_{adv} = \Sigma v_r T \frac{dS}{dR} = -\frac{\dot{M}}{2\pi R}T\frac{dS}{dR} = \frac{\dot{M}}{2\pi R^2}\frac{P}{\rho}\xi, \qquad (4)$$

where $\Sigma = 2H\rho$ is the surface density of the disk, $S$ is the specific entropy, and $\xi = -[(4-3\beta)/(\Gamma_3 - 1)](d\ln T/d\ln R) + (4-3\beta)(d\ln\Sigma/d\ln R)$. Here, $\beta = P_g/P$, $\Gamma_3 = 1 + (4-3\beta)(\gamma-1)/[\beta + 12(\gamma-1)(\beta-1)]$, and $\gamma$ is the ratio of specific heats. For optically thin disks $\beta = 1$.

Using equations (2-4), the energy equation, $Q_+ = Q_{adv} + Q_{brem}$, can be written as

$$\xi \dot{m}^2 - 0.361 r^{1/2} g^2 q^{-1} (\alpha \Sigma) \dot{m} + 3.462 \times 10^{-6} \alpha^{-2} r^2 f^{-1} g (\alpha \Sigma)^3 = 0, \tag{5}$$

where $q = 1 - R_G/R$, $r = R/R_G$, $\dot{m} = \dot{M}/\dot{M}_E$. Here $\dot{M}_E = 4\pi GM/(c\kappa_{es})$ is the Eddington accretion rate defined as the Eddington luminosity devided by $c^2$. In deriving equation (5) we have used the standard equations:

$$\nu = \frac{2}{3} \alpha C_s H, \tag{6}$$

$$(H/R)^2 = (\sqrt{2}/\kappa_{es}) r^{-1/2} \dot{m} (\alpha \Sigma)^{-1} f g^{-1} q = 2r \frac{P}{\rho c^2} q^2, \tag{7}$$

$$\dot{M} = -2\pi R v_r \Sigma. \tag{8}$$

Equation (5) gives the equilibrium relation between $\dot{m}$ and $\Sigma$ for a given $\alpha$ and $\xi$ for any radius. Since it is a quadratic in $\dot{m}$ for a given $\Sigma$ one expects at most two possible values of $\dot{m}$. It is possible to show numerically that a critical $\Sigma$ value exists above which there is no solution to $\dot{m}$. This corresponds to a single solution $\dot{m}_{max}$:

$$\dot{m}_{max} = 1.7 \times 10^3 r^{-1/2} \alpha^2 f g^5 q^{-3} \xi^{-2}. \tag{9}$$

One notes that it scales to $\alpha^2$. The relation of $\dot{m}_{max}(r)$ for a given $\alpha$ and $\xi$ is shown in Figure 2. One can see that for $\alpha \lesssim 0.1$, $\dot{m}_{max}$ corresponds to a very low mass accretion rate. Furthermore, even for $\dot{m} < \dot{m}_{max}$, only a fraction of the disk can be optically thin; and the lower the mass accretion rates, the wider the possible optically thin region of the accretion disk.

The solution of equation (5) on the $(\dot{M}, \alpha \Sigma)$–plane is shown in Figure 3a and b, together with the S-shaped sequence corresponding to the optically thick equilibria. Disk parameters of $M/M_\odot = 10$, $r = 5$, and $\alpha = 0.1$ and $0.01$ were used. We also assumed $\xi$ is unity. For the case of small viscosity, i.e., $\alpha \lesssim 0.1 \sim 0.2$, the optically thin equilibria form a sequence disjoined from that formed by the optically thick equilibria. The optically thick sequence consists of three branches: (1) gas pressure supported, radiatively cooled one, which is both thermally and viscously stable, (2) radiation pressure supported, radiatively cooled one, which is both thermally and viscously unstable, (3) radiation pressure supported, advectively cooled one, which is both thermally and viscously stable. The first two branches are described by the standard Shakura-Sunyaev solution, the third one is given by the slim disk solution. The optically thin sequence consists of two branches: (4) gas pressure supported, radiativelly cooled one, which is viscously stable but thermally unstable, and (5) gas pressure supported, advectively cooled one, which is both thermally and viscously stable. The fact that it is viscously stable should be obvious from its positive slope. Thermal stability can be infered from the fact that above that line the heating rate is less than the cooling rate but it follows from a criterion properly modified to include non-locality of the advection (Abramowicz et al 1994). Similar three stable branches are found in the boundary layer numerical calculations of Narayan & Popham (1993)

The explicit analytical form of this stable equilibrium was not known before. One notes from equation (5) that it corresponds to a relation

$$\dot{m} \approx 0.361 \xi^{-1} r^{1/2} g^2 q^{-1} \alpha \Sigma. \tag{10}$$

A relation of the form $\dot{m} = f(r)\alpha\Sigma$ is characteristic of advection dominated $\alpha$-disk models as first noticed by Abramowicz et al (1988; see their eq. [29]). It is also a property of advective dominated sub-Keplerian self-similar flows found by Narayan & Yi (1994), which is $\dot{m} \approx 0.1 r^{1/2} \alpha \Sigma$ for their solutions with small $\alpha$ and $\gamma = 5/3$. In general the form of the function $f$ in the $\dot{m} = f\alpha\Sigma$ relation depends on the equation of state and $v_r$.

The parameter of $\xi$ can be estimated approximately for the advection dominated branch by assuming the correction terms, $f$, $g$ and $q$ to be unity. Combining equations (7), (10), and equation of state, we have $T \propto r^{-1}$ and $\Sigma \propto r^{-1/2}$. Hence, $\xi = 1/(\gamma-1)-1/2$, which is 1 and 5/2 for $\gamma = 5/3$ and 4/3 respecively.

One should keep in mind that equilibria discussed in this paper were analysed only at a given radius. To get correct radial dependences one should solve the full set of conservation equations in a similar way as it was done for the optically thick slim disk models. Detailed global properties of advection dominated accretion disks will be studied in a future work.

7MAA and JPL thank the Theoretical Physics Institute of the Technion, Haifa, Israel for its kind hospitality in April and May 1994 when part of this work was performed. XC thanks Prof. Ron Taam for discussions on the physics of accretion and advection. We thank an anonymous referee for useful suggestions.

# FIGURE CAPTIONS

**Figure 1.—** (a) Thermal equilibra from Kusunose & Mineshige (1992), $\alpha = 0.1$. The solid, dashed, long dashed, and dash-dotted lines are for solutions with $e^+e^-$ pairs, without pairs, one-temperature, and optically thick disks respectively. Advection is dominant above the heavy dotted line. The heavy dashed line correspond to $H/R = 1$. (b) Thermal equilibra from Luo & Liang (1994). The solid triangles, open triangles and stars are for solutions with $\alpha$ of 1, 0.1, and 0.01 respectively. The solid, dotted, and dashed lines correspond to advection domination limits for the same sequential of $\alpha$.

**Figure 2.—** The variation of $\dot{m}_{\max}$ with respect to the radius of the disk. Here $\xi = 1$ is assumed. The solid and dotted lines are for $\alpha = 0.1$ and 0.01 respectively. Note that for $\alpha \lesssim 0.1$, the disk can be optically thin only at very low mass accretion rates.

**Figure 3.—** (a) Thermal equilibria for optically thick (the right solid S-shaped line) and optically thin (the left solid line) accretion disks. The upper branches represent advection dominated solutions. Configuration above the dotted lines $\tau = 1$ are optically thin, where $\tau$ is the effective optical depth calculated by assuming that the pressure is dominated either by radiation (the upper one) or by gas (the lower one). The parameters assumed here are $M/M_\odot = 10$, $r = 5$, $\alpha = 0.1$, and $\xi = 1$. (b) The same as (a) except for $\alpha = 0.01$.